\documentclass{aa}
\usepackage{graphics}

\def\kms    {\ifmmode{{\rm ~km~s}^{-1}}\else{~km~s$^{-1}$}\fi}

\def\arcmper  {\ifmmode \rlap.{' }\else $\rlap{.}' $\fi}
\def\arcsec   {\mbox{$^{\prime\prime}$}}
\def\arcsper  {\ifmmode \rlap.{'' }\else $\rlap{.}'' $\fi}
\def\arcsgper  {\ifmmode \rlap.^{s }\else $\rlap{.}^s $\fi}

\def\deg      {\ifmmode^\circ\else$^\circ$\fi}     

\def\sper     {\ifmmode \rlap.^{s}\else $\rlap{.}^s$\fi}

\def\>           {$>$}
\def\<           {$<$}

\def\kms    {\ifmmode{{\rm ~km~s}^{-1}}\else{~km~s$^{-1}$}\fi}

\begin{document}


\title{Revised positions for CIG galaxies}
\author{S. Leon\inst{1}, L. Verdes-Montenegro \inst{1}}
\institute{Instituto de Astrofisica de Andalucia, Ap. 3004, Granada 18080, SPAIN
stephane@iaa.es,lourdes@iaa.es}
 
\offprints{S. Leon}
\date{Received ; accepted }
\titlerunning{Revised positions for the CIG galaxies}

\authorrunning{ .}

\maketitle
\begin{abstract}
We present revised positions for the 1051
galaxies belonging to the Karachentseva Catalog of Isolated
Galaxies (CIG). New positions were calculated by
applying SExtractor  
to the Digitized Sky Survey CIG fields 
with a spatial resolution of 1\arcsper 2.
We visually checked the results and
for 118 galaxies had to recompute the assigned positions
due to complex morphologies (e.g. distorted isophotes, 
undefined nuclei, knotty galaxies) or the presence of bright stars.
We found differences between older and newer positions
of up to 38\arcsec\ with a mean value of 2\arcsper 96 
relative to SIMBAD and up to 
38\arcsec\ and 2\arcsper 42 respectively relative to UZC.
Based on star positions from the APM catalog we determined that the DSS
astrometry of five CIG fields has a mean offset in ($\alpha$,$\delta$) of 
(-0\arcsper 90,0\arcsper 93) with a dispersion of 0\arcsper 4.
These results have been confirmed using the 2MASS All-Sky Catalog of
Point Sources.
The intrinsic errors of our method combined with the astrometric ones 
are of the order of 0\arcsper 5.
\keywords{galaxies, surveys, astrometry}
\end{abstract}

\section{Introduction}

The evolutionary history of galaxies
 can be strongly influenced by the environment. Therefore a
definition of ``isolated galaxy'' is needed before one can properly
assess the history and properties of interacting or peculiar ones.
This motivated us to assemble a well-defined and statistically
significant sample of isolated galaxies to serve as a comparison
template in the study of galaxies in denser environments
(Verdes-Montenegro et al 2001, 2002; Lisenfeld et al 2002ab).  
We will make public all obtained data for this sample via the web at 
http://www.iaa.csic.es/AMIGA.html where the results from this paper
can be already retrieved.
Our working
sample is drawn from the Catalog of Isolated Galaxies (CIG, also referred
as K73 in SIMBAD and KIG in NED databases)
 which originally contained
n=1051 galaxies  and was selected on the basis of the distance to the nearest
similarly sized galaxies (no other galaxy within 4 times its diameter
and within a distance of 20 times their size; Karachentseva 1973; 
see also Sulentic 1989).  Later the sample was reduced to n=893 galaxies
(Karachentseva 1980), the rest showing less strict degrees of isolation. 
During preparatory
work on the survey we  searched  for CIG positions in the SIMBAD
database and noticed shifts with respect to the central positions
of the galaxies, reaching in some cases up to several tens of 
arcseconds.  This fact not only prevents accurate pointings for 
reduced fields of view, but also makes  cross-identifications with 
other available catalogs more difficult. We then searched   the 
Updated Zwicky Catalogue (UZC; Falco et al. 1999)
whose accuracy  peaks at 1\arcsec\
with a width of 1\arcsper 45, as estimated after matching 
with the FIRST 1.4 GHz catalog (White et al. 1997).
Still in a preliminary exploration mode
we found CIGs with 
positional errors larger than 10\arcsec\ in the UZC  (see Sect. 2).
This motivated us to revise the positions of the
entire CIG in a systematic way. 
  We decided to provide
positions for the entire catalog without consideration of any 
isolation criterion.  In Sect. 2 we explain our method and compare
our results with other surveys and in Sect. 3 we give notes for
individual galaxies. Our conclusions are presented in Sect. 4.

\section{New CIG positions}

 We obtained  Digitized Sky Survey (DSS) red images
of all CIG galaxies in J2000 coordinates, with pixel
sizes of  1\arcsper 2 and 
6\arcmin $\times$ 6\arcmin\, fields. The images were analyzed 
in an automatic procedure  using the SExtractor software 
 (Bertin \& Arnouts 1996)  for
sources brighter than 4$\sigma$ the background level. Those
closer to the original CIG position were automatically selected
but later revised visually in order to confirm that 
we had targeted the right galaxy. 
The internal error of the position fit is better than 0\arcsper 05.

Once the automatic process was done, a visual check of the SExtracted 
positions was performed for all the galaxies. 
In 118 cases we had to recompute the assigned positions 
due to one of the following problems:
1)   4$\sigma$ isophote not 
indicative of the nucleus positions (e.g. distorted isophotes),
2) ill-defined (low contrast) nucleus,
3)  presence of brighter off-centered regions 
(e.g. irregular/clumpy galaxies), a star superposed on 
 the galaxy, 
and 4) in 3 cases the galaxy was in fact 
a misclassified globular cluster or a dwarf galaxy of the Local Group (CIG 388, 781 and 802,
see Section 3)
and we have excluded them from the statistics of our study.
Problematic positions have been recomputed in a second iteration following  one of 
two procedures:
(a) changing the SExtractor parameters ($\sigma$ threshold and background 
rms determination) in order to select the most regular isophote or the brightest 
central region according to the galaxy morphology/central brightness distribution; 
(b) visually in the five cases  where a star was interfering
with the galaxy image
(see Sect. 3). 

\begin{figure*}{th}
\resizebox{16cm}{8cm}{\includegraphics{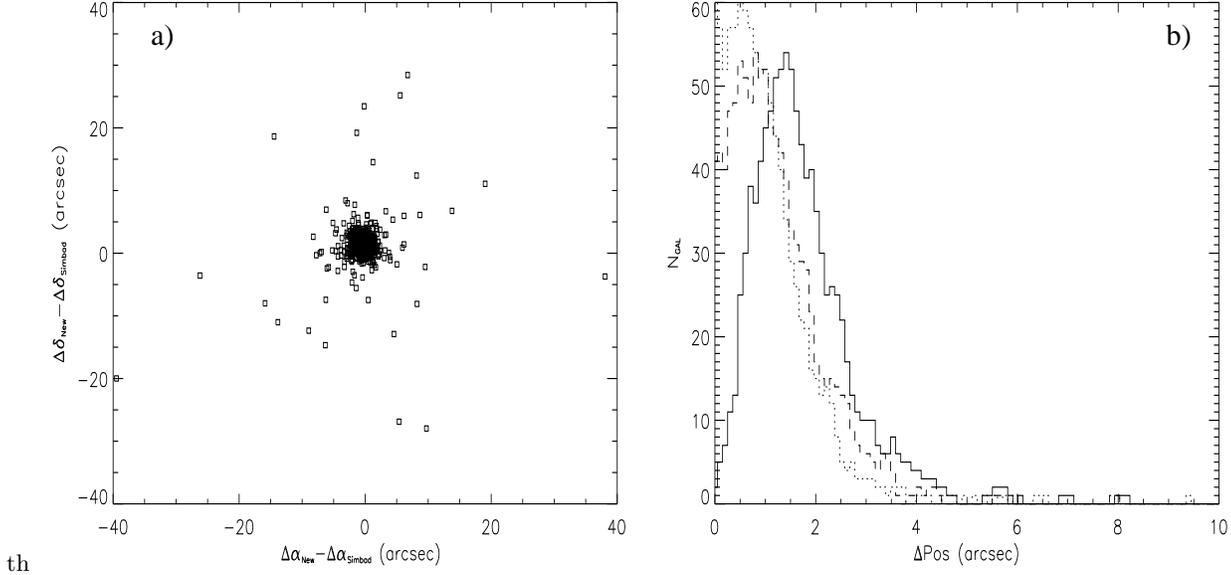}}
\caption{a) Differences between our
measured positions and those retrieved from SIMBAD for the CIG galaxies. 
 b) Histogram of the difference between the new
positions and the SIMBAD positions for the $\alpha$ (dotted line),
$\delta$ (dashed line) coordinates and the total distance (solid line) in 
arcsec. The plotted range is restricted to 10\arcsec\ for clarity of the plot.
 }
\label{fig_diff_simbad}
\end{figure*}

We calculated the differences found 
in $\alpha$ and $\delta$ between our estimated positions and 
those obtained from SIMBAD (see Figure  \ref{fig_diff_simbad}).
The difference G in $(\alpha,\delta)$ position  is fitted by a 2D Gaussian as follows:

\begin{equation}
G(\alpha,\delta)=C+K*e^{-\frac{1}{2}\left[ \left ( \frac{\alpha-\alpha_0}{\sigma_\alpha} 
\right )^2+ 
\left ( \frac{\delta-\delta_0}{\sigma_\delta} \right )^2 \right ]
}
\end{equation}

 The dispersion in  position differences relative to SIMBAD is
 isotropic with
 $(\sigma_\alpha,\sigma_\delta)=(0\arcsper 75,0\arcsper 78)$, as can be
 seen in Figure \ref{fig_diff_simbad}a.  The histograms of the
 $\alpha$, $\delta$ and total difference are shown in Figure
 \ref{fig_diff_simbad}b: the total difference has a dispersion of
 2\arcsper 96 with a maximum of 38\arcsec\ between the two positions 
for CIG 239.
 We also find a shift between our estimates and SIMBAD positions of
 $(\alpha_0,\delta_0)= (-0\arcsper 76,1\arcsper 01)$. We discuss
below error sources including the possible origin of the found shift.

\begin{figure}
\resizebox{8cm}{8cm}{\includegraphics{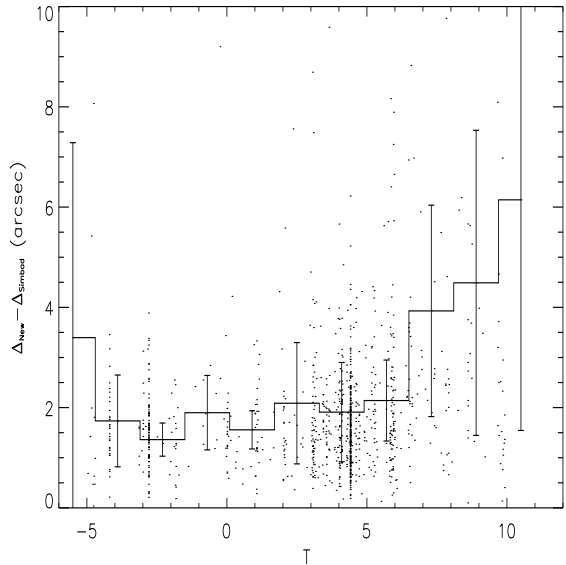}}
\caption{Differences between our
measured positions and those retrieved from SIMBAD 
for the CIG galaxies as a function of 
 the morphological type T. A binned average (solid line) as well as 
1$\sigma$ dispersion (error bars) are also shown.}
\label{fig_diff_morfo}
\end{figure}

\begin{figure}
\resizebox{8cm}{8cm}{\includegraphics{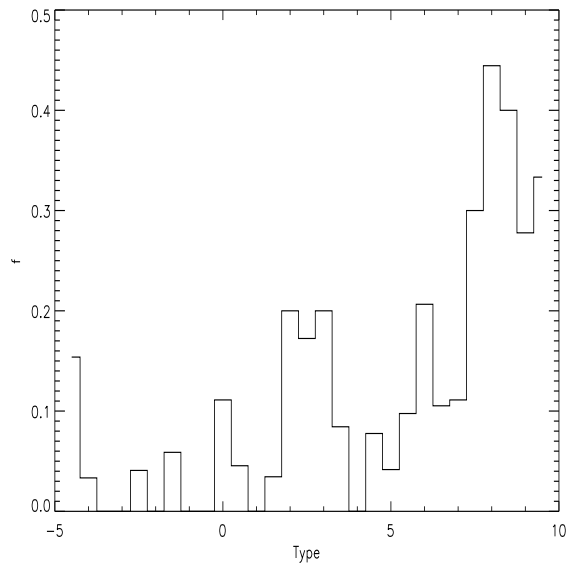}}
\caption{Fraction of galaxies of a given morphological
type that we had to reprocess due to the problems described in Sect. 2. }
\label{fig_reproc}
\end{figure}

 The internal positional error of
 SExtractor is $<$ 0\arcsper 05.  
The two larger sources of errors are associated with the optical morphology of 
the galaxies and the DSS astrometry.  The 
dispersion introduced by the smaller bulge galaxies 
is evident when we compare (Figure \ref{fig_diff_morfo}) the position differences 
between SIMBAD and this work  
against the
morphological type T of the galaxies (de Vaucouleurs et al 1991, 
with $-5 \le T < 0$ for E/S0 galaxies, $0 \le T<9$ for the spiral galaxies 
 and $T \ge 9$ for the
irregular ones). The differences are clearly increasing 
from $\sim 1\arcsper 5$ up to $\sim 4\arcsper 5$
from the early type galaxies with a large bulge towards the irregular
galaxies. The two extreme
values are highly dispersed mainly because of the low number of
galaxies falling in the bin. In fact the majority of the 
118 problematic galaxies
that we had to reprocess have late morphological types (Figure  \ref{fig_reproc}).
The associated errors are difficult to estimate, but affect
only 10\% of the total sample.

\begin{figure}
\resizebox{8cm}{8cm}{\includegraphics{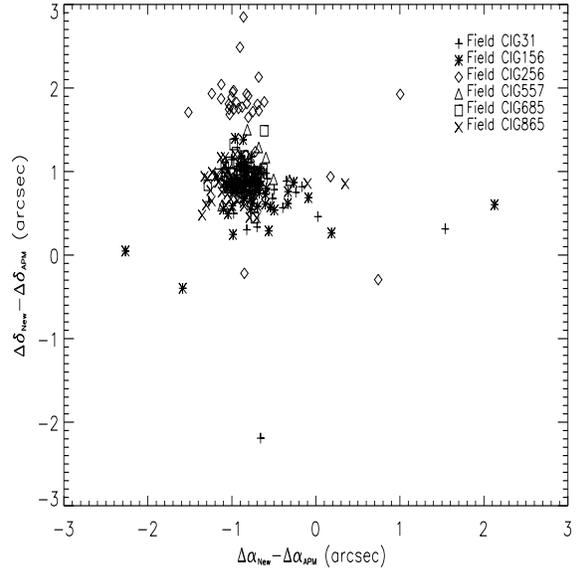}}
\caption{Differences between our
measured positions and those retrieved from the APM catalogue  for the sources extracted
in the six CIG fields indicated in the upper right corner of the plot. }
\label{fig_diff_apm}
\end{figure}

\begin{figure}
\resizebox{8cm}{8cm}{\includegraphics{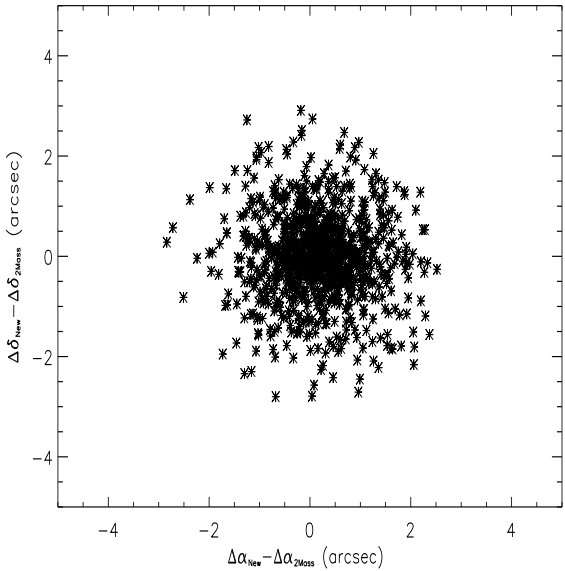}}
\caption{Differences between our CIG
measured positions and those retrieved from the 2MASS  All-Sky Catalog of Point Sources
 after correcting for a mean offset of 
(-0\arcsper 90,0\arcsper 93)
for  the ($\alpha,\delta$) coordinates (see Sect. 2). 
}
\label{fig_2mass}
\end{figure}

In order to analyze DSS astrometry we
 selected 6 different CIG fields spread over the
sky and compared the positions of 762 stars extracted from these fields
 with positions from the APM catalogue (Automated Photographic Measuring,
Maddox et al 1990)
 which was calibrated using more than 200 PPM
(Catalogue of Positions and Proper Motions, Roser \& Bastian
1991) 
astrometric standards per Palomar field, reaching an accuracy of
0\arcsper 5  (e.g. Arnouts et al. 1999). The differences in
coordinates are shown in Figure \ref{fig_diff_apm} for the 6 CIG
fields. A mean offset, similar to the one found
previously, is still present,  with the exception of the CIG 256 
field which has an extra offset in $\delta$ of $\sim$0\arcsper 7. We have fitted the mean offset
($\alpha,\delta$) as previously, excluding the CIG 256 field: a mean
value of (-0\arcsper 90,0\arcsper 93) is found with a dispersion of 0\arcsper 4
.  This value is consistent 
with the offset we found between SIMBAD and DSS extracted
positions, suggesting that it was 
introduced by the DSS astrometry. 
We performed a complementary  
check using the 2MASS All-Sky Catalog of Point Sources
via VizieR Service at CDS and 
found 931 sources closer than  3\arcsec\ 
 to the CIG corrected positions. 
The differences between our CIG measured positions corrected
by the mean offset of (-0\arcsper 90,0\arcsper 93) and 
those retrieved from the 2MASS catalog for these sources 
(Figure \ref{fig_2mass}) have a mean value of 
($\alpha,\delta$) =  (-0\arcsper 16,-0\arcsper 00)
with a dispersion of (0\arcsper82,0\arcsper88). 
This comparison hence supports the applied astrometric
correction and 
 the mean error due to astrometry based on this
analysis is estimated to be
at the  0\arcsper 5 level.

\begin{figure}
\resizebox{8cm}{8cm}{\includegraphics{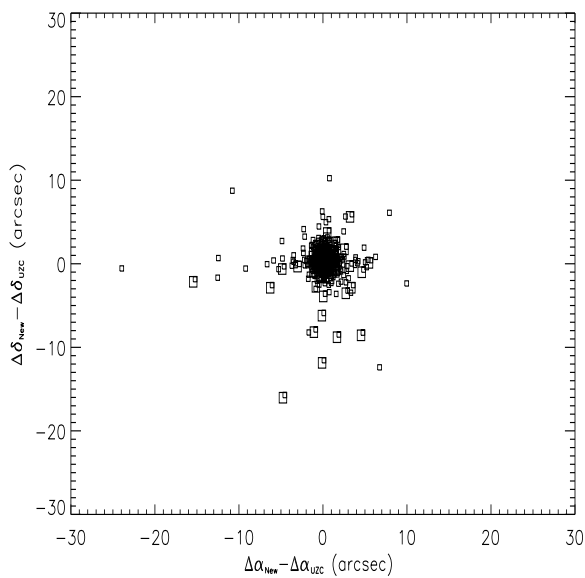}}
\caption{Differences between our CIG
measured positions and those retrieved from the UZC 
 after correcting for a mean offset of 
(-0\arcsper 90,0\arcsper 93)
for  the ($\alpha,\delta$) coordinates (see Sect. 2). 
 We have excluded two points with larger 
differences in benefit of a clearer plot. Large squares 
correspond to visually inspected galaxies.}
\label{fig_diff_uzc}
\end{figure}

\begin{figure*}
\resizebox{16cm}{8cm}{\includegraphics{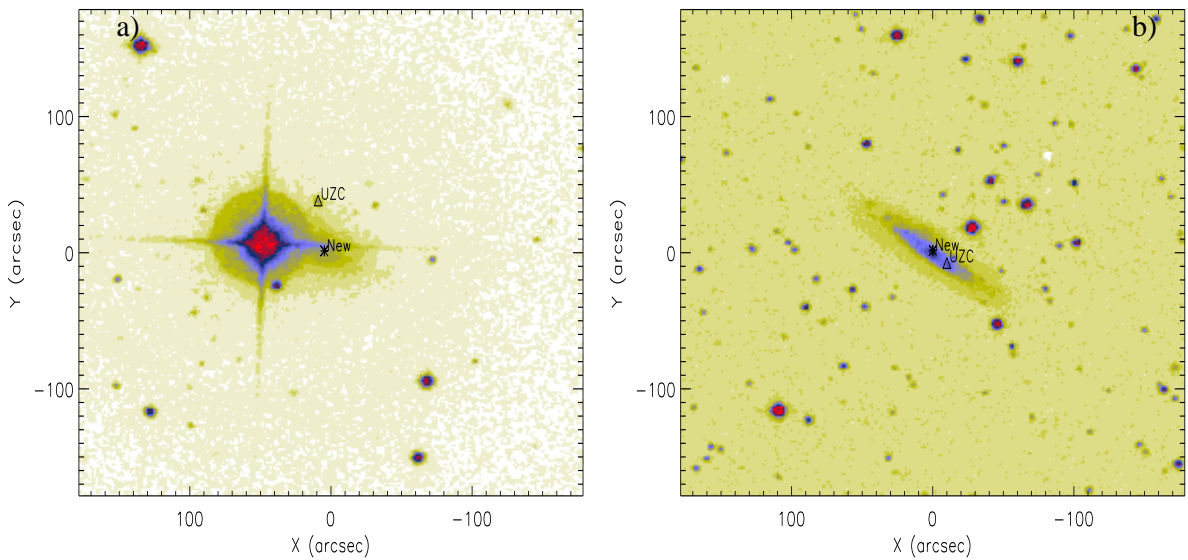}}
\caption{DSS image of the (a) CIG 402 field, where a bright star
is superposed on the galaxy and (b) CIG 828 field.  The stars  indicate our newly
calculated positions, whereas the triangles correspond to the UZC
position.}
\label{fig_diff_cig402}
\end{figure*}

After  this correction  we  compared the new positions
with the  749 galaxies in common with the UZC. 
  Figure \ref{fig_diff_uzc} shows the difference in
($\alpha,\delta$) coordinates between our revised positions and the
UZC.  The dispersion in both coordinates $\alpha$ and $\delta$
are small (3\arcsper 2,2\arcsper 5) with a total dispersion of 20\arcsper 42.
 Nevertheless the mean offset in the
($\alpha,\delta$) is only (-0\arcsper 17,0\arcsper 06), negligible for
the accuracy targeted.  The maximum difference between our new
positions and the UZC reaches 38\arcsec\ for CIG 402 where a
very bright star is interfering with the galaxy. 
In Figure \ref{fig_diff_cig402} we illustrate the differences
found between our measured positions and the ones given in UZC 
with two examples.

The new positions are given in Table 1 for the first 10 galaxies and the 
rest are available in electronic form at the CDS via anonymous ftp to 
cdsarc.u-strasbg.fr (130.79.128.5) or in our public database
available from http://www.iaa.csic.es/AMIGA.html.
Column (1) gives the CIG identification, Column (2) alpha and
delta in B1950 coordinates, Column (3) alpha and 
delta in J2000 coordinates and Column (4)  the $\sigma$ threshold for the source detection
 when  different from the  4$\sigma$ value chosen by default.  V indicates a position estimated visually while
E identifies the three excluded objects.

\begin{table*}
      \caption{CIG positions determined from DSS images}
\begin{tabular}{rrrrrrrrrrrrrc}
CIG  & \multicolumn{3}{c}{$\alpha$(1950)}&
\multicolumn{3}{c}{$\delta$(1950)}&
\multicolumn{3}{c}{$\alpha$(2000)}&
\multicolumn{3}{c}{$\delta$(2000)}&
$\sigma$ threshold$^{1}$\\
    1 &    0& 00& 31.92 &  -2& 11& 33.8&     0& 03&  5.63&   -1& 54& 51.6 &   30 \\
    2 &    0& 00& 46.30 &  29& 31&  8.0&     0& 03& 20.37&   29& 47& 50.2 &  \\
    3 &    0& 00& 47.77 &  30& 30& 13.4&     0& 03& 21.86&   30& 46& 55.7 &  \\
    4 &    0& 01& 24.75 &  20& 28& 26.5&     0& 03& 58.77&   20& 45&  8.7 &    30 \\
    5 &    0& 05& 19.86 &  20& 08&  2.8&     0& 07& 54.30&   20& 24& 44.6 & \\
    6 &    0& 06& 20.00 &  23& 32& 21.1&     0& 08& 54.70&   23& 49&  2.8 &   29 \\
    7 &    0& 08& 32.54 &   2& 23& 59.5&     0& 11&  6.39&    2& 40& 40.8 & \\
    8 &    0& 09& 34.64 &  11& 46&  2.9&     0& 12&  9.04&   12& 02& 44.0 & \\
    9 &    0& 10&  4.23 &   5& 13& 38.2&     0& 12& 38.26&    5& 30& 19.2 & \\
   10 &    0& 10& 24.24 &  38& 58&  4.0&     0& 13&  0.74&   39& 14& 45.0 & \\
\end{tabular}
\begin{list}{}{}
\item[$^{\rm 1}$] The $\sigma$ threshold used in SExtractor 
for the source detection is indicated
 when  different from the  4$\sigma$ value chosen by default.
A V appears when the positions was visually estimated and an E when
the object was excluded from our study as explained in Sect. 2.
\end{list}
\end{table*}

\section{Notes on individual galaxies}

CIG~63  - Center undefined, eccentric bright peak was chosen.\\
CIG~190 - Center obtained from a 5$\sigma$ isophote since the galaxy appears faint 
with an ill-defined nucleus. \\
CIG~235 -   Center obtained from a 20$\sigma$ isophote due to the knotty morphology
of the galaxy.\\
CIG~239 -  Center undefined,  fitted with a 25$\sigma$ threshold
 on the optical peak of the bright eastern  feature.\\
CIG~261 -  Center fitted with a 30$\sigma$ threshold on the offcentered bright peak.\\
CIG~388 - This is a Local Group member (Sextans B).\\
CIG~402  - Center defined visually  because of a  bright star close to the object.\\
CIG~523  - Center fitted with a 20$\sigma$ threshold on the brightest central cluster.\\
CIG~530  - Diffuse galaxy, undefined center.\\
CIG~569  - Clumpy galaxy, center fitted with a   25$\sigma$ threshold  on the central bright cluster.\\
CIG~621  - Offcentered nucleus.\\
CIG~649  - Center defined visually  because of a  bright star close to the object.\\
CIG~781  - Not a galaxy but the globular cluster Pal 5.\\
CIG~802 -  This is a Local Group member (Draco).\\
CIG~810  - Center defined visually  because of a star overlapping
with  the center of this edge-on galaxy.\\
CIG~853  - Center fitted with a 40$\sigma$ threshold  on the brightest  region  on the west side.\\
CIG~883  - Center defined visually  because of a star overlapped with
 this  galaxy. \\
CIG~928  - Center fitted with a 20$\sigma$ threshold on the  northern object close since
the Southern object appears to be a star.\\
CIG~947  - Offcentered nucleus, fit to the 22$\sigma$ isophote.\\
CIG~959  - Center defined visually  because of a star overlapped with
 this  galaxy. \\
CIG~967  - Extended bright center, fit to the 20$\sigma$ isophote.\\
CIG~977  - Center fitted with a 10$\sigma$ threshold on the  Northern bright region. \\
CIG~1036 - Center fitted with a 25$\sigma$ threshold on bright region of this distorted
 galaxy.\\

\section{Conclusions}

The CIG galaxies' positions have been recomputed for the whole sample (1051
galaxies) using an automatic procedure of source extraction on the
DSS images plus a visual check.  The comparison with previous positions
from the SIMBAD database shows an average offset of
$(-0\arcsper 76,1\arcsper 01)$ for the $(\alpha,\delta)$ coordinates with
a dispersion in the difference in position of 3\arcsper 0. A
check with the APM positions of stars in different CIG fields,
and with sources in common with the 2MASS All-Sky Catalog of Point Sources
 shows
that this offset is due to the DSS astrometry. This mean offset has
been
corrected in our derived positions.  The comparison with the UZC
arcsec accuracy positions on 749 overlapping galaxies shows a
negligible mean offset, with a dispersion of $\sim$2\arcsper 4.  
Taking into account the internal error
from SExtractor and the error from astrometry, 
these new positions are correct within an error
of $\sim$ 0\arcsper 5 for 90\% of the CIG galaxies. For the 
rest of galaxies the errors are linked with  late 
morphological types that required an interactive reprocessing
of the data.

\begin{acknowledgements}
SL and LVM are partially supported by spanish MCyT  Grant  
AYA 2002-03338. S. L. is supported
by a Marie Curie Individual Fellowship contract HPMF-CT-20002-01734
from the European Union.  Based on photographic data obtained using
the UK Schmidt Telescope.  The UK Schmidt Telescope was operated by
the Royal Observatory Edinburgh, with funding from the UK Science and
Engineering Research Council, until 1988 June, and thereafter by the
Anglo-Australian Observatory. Original plate material is copyright ©
the Royal Observatory Edinburgh and the Anglo-Australian Observatory.
The plates were processed into the present compressed digital form
with their permission.  The Digitized Sky Survey was produced at the
Space Telescope Science Institute under US Government grant NAG
W-2166. This research has made use of the SIMBAD database, operated at
CDS, Strasbourg, France. 
This publication makes use of data products from the Two Micron All
Sky Survey, which is a joint project of the University of
Massachusetts and the Infrared Processing and Analysis
Center/California Institute of Technology, funded by the National
Aeronautics and Space Administration and the National Science
Foundation. We acknowledge useful comments by
J. Guibert. We would like 
to acknowledge the anonymous referee for a  careful 
reading of the manuscript.

\end{acknowledgements}

\end{document}